%% file: IndustrialOuroboros.tex
\documentclass[letterpaper,twocolumn,10pt]{article}
\usepackage{usenix-2020-09}
\usepackage{float}
\usepackage{url}

\usepackage{tikz}
\usepackage{amsmath}

\usepackage{filecontents}
\begin{document}

\date{}

\title{\Large \bf Industrial Ouroboros: Deep Lateral Movement via Living Off the Plant}

\author{
{\rm Richard Derbyshire}\\
Orange Cyberdefense\\
ric.derbyshire@orangecyberdefense.com
}

\maketitle

\begin{abstract}
Lateral movement is a tactic that adversaries employ most frequently in enterprise IT environments to traverse between assets.
In operational technology (OT) environments, however, few methods exist for lateral movement between domain-specific devices, particularly programmable logic controllers (PLCs).
Existing techniques often rely on complex chains of vulnerabilities, which are noisy and can be patched. 
This paper describes the first PLC-centric lateral movement technique that relies exclusively on the native functionality of the victim environment. 
This OT-specific form of `living off the land' is herein distinguished as `living off the plant' (LOTP).
The described technique also facilitates escape from IP networks onto legacy serial networks via dual-homed PLCs.
Furthermore, this technique is covert, leveraging common network communication functions that are challenging to detect.
This serves as a reminder of the risks posed by LOTP techniques within OT, highlighting the need for a fundamental reconsideration of traditional OT defensive practices.
\end{abstract}

\input{1Introduction/1introduction}
\input{2Related/2related}

\input{3Background/3background}
\input{4Implement/4implement}
\input{5Conclusion/5conclusion}

\bibliographystyle{plain}
\bibliography{bibfile}

\end{document}

%% file: 1Introduction/1introduction.tex
\section{Introduction}
\label{intro}

\begin{figure}[]
    \centering
    \includegraphics[width=\linewidth]{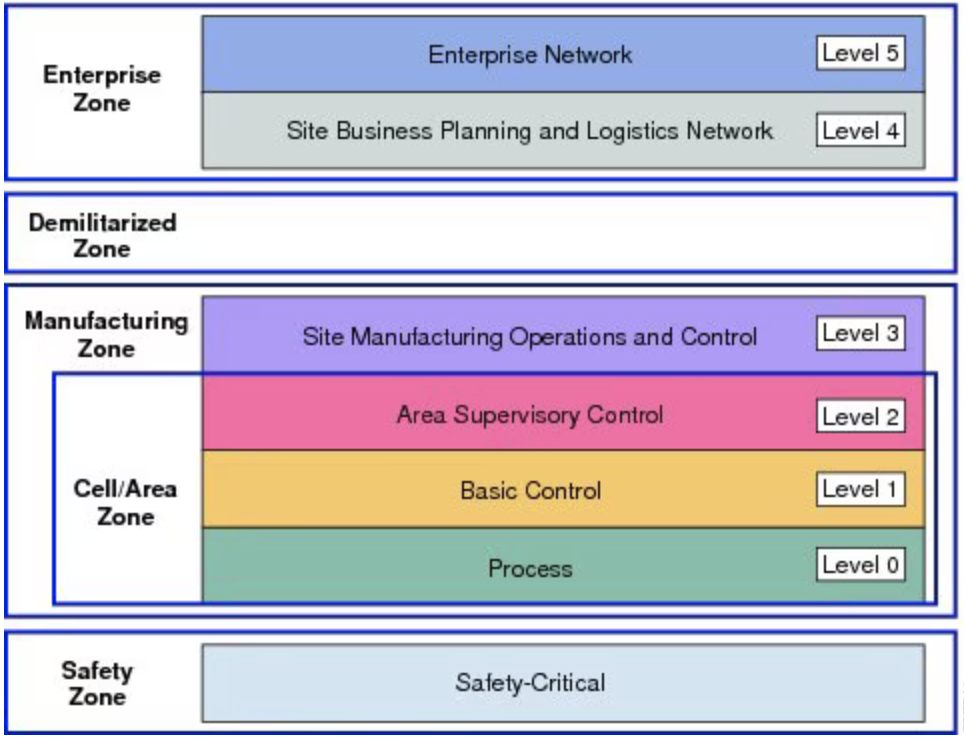}
    \vspace{-15pt}
    \caption{Purdue Enterprise Reference Architecture~\cite{CiscoPERA}}
	\label{fig:purdue}
    \vspace{-10pt}
\end{figure}

Operational technology (OT) powers much of today's critical national infrastructure, including water treatment and energy distribution~\cite{Stouffer2023}.
Unlike their information technology (IT) counterparts, OT environments consist of assets that facilitate interaction with and automation of physical operational processes. 
Figure \ref{fig:purdue} depicts the Purdue Enterprise Reference Architecture (i.e., the Purdue Model), which provides a high-level view of a typical OT deployment and its interface with IT environments.
The lower the level, the closer its assets are to the operational process and physical world, with the demilitarised zone (DMZ) marking the boundary between OT and IT.
At the heart of an OT environment, programmable logic controllers (PLCs) connect to operational processes through sensors and actuators, enabling monitoring, control, and automation via programmed logic and operator input.\\

Adversaries employ a range of tactics to achieve specific objectives during cyber attacks, applicable across both IT \cite{MitreAttack} and OT \cite{MitreICSAttack} environments.
Of those tactics, lateral movement \cite{MitreICSLateral} is used to traverse between assets and across the victim's environment.
It is used for a variety of reasons, including moving to assets of interest, maintaining persistence, evading detection, or even indirectly elevating privileges.
Many of the techniques used to achieve lateral movement are IT-focused, with emphasis put on moving between IT assets, or from IT to OT via the DMZ.
However, techniques for lateral movement between OT assets in the lower levels of the Purdue Model are seldom achieved, and due to their rarity are considered valuable \cite{Derbyshire2022}.
This is especially true for operational PLCs in level 1, and safety PLCs in the safety zone. 

Living off the land (LOTL) is an approach where adversaries leverage legitimate functionality within the victim's environment.
Rather than relying on external exploits or malware, adversaries use native software, tools, and features to achieve their objectives during a cyber attack.
This type of approach can be amplified further by poor practice such as software misconfigurations, weak architectures, or exposed file shares.
By foregoing external tools in favour of those native to the victim's environment, LOTL approaches are inherently more difficult to detect and therefore offer greater stealth. 
This covert nature allows adversaries to operate for longer periods without triggering traditional security alerts.
When applied to an OT context, LOTL reveals a unique and underexplored attack surface that can be used to target and manipulate operational processes in ways that traditional cyber attack techniques cannot \cite{Derbyshire2024,Greenpcaad2021}.
Unlike in IT environments, where LOTL often focuses on administrative tools or user software, in OT it exploits the very mechanisms that control the physical process.
To capture this distinction with clarity, this paper refers to the OT variant as `living off the plant' (LOTP).

This paper, therefore, introduces a novel technique that takes advantage of LOTP to achieve lateral movement between PLCs, catalysing multiple novel capabilities.
In particular, this technique can exploit both site-to-site communications between PLCs over wide area networks (WAN) and native serial communications, enabling movement beyond traditional network boundaries. 
As a result, it has the potential to impact operations across large geographic areas and, in some cases, may even reach devices such as safety PLCs that were previously considered unreachable.

The remainder of this paper is structured as follows: 
Section~\ref{related} explores related work in the field of lateral movement techniques and LOTP methods in OT. 
Section~\ref{background} first outlines PLC fundamentals by describing their memory and programming architecture, and then introduces PLC-centric lateral movement threat models relevant to this technique.
Section~\ref{practical} presents proof-of-concepts, illustrating the realisation of each threat model at a practical level. 
Section~\ref{conclude} concludes the paper by articulating the necessity for further research into both lateral movement techniques and LOTP, as well as their implications for current security practices.

%% file: 2Related/2related.tex
\section{Related Work}
\label{related}

Historic attacks targeting OT environments have demonstrated a range of offensive techniques across all levels of the Purdue Model~\cite{Miller2021, Derbyshire2018}. 
However, lateral movement has been most notably utilised by adversaries to traverse from the victim's IT environment to their OT environment.
For example, the 2015 attack on Ukraine's electricity distribution network leveraged IT helpdesk software to reach into the OT environment, releasing electricity breakers through the direct control of operator stations~\cite{Ukraine2015}. 
Lateral movement within OT environments tends to treat the lower levels of the Purdue Model as a destination or target, rather than a pivot point. 
The TRITON malware detected in a Saudi Arabian petrochemical plant in 2017 acts as one such example. 
It was launched from a trusted Windows-based Triconex engineering workstation, towards Tricon controllers~\cite{Triton2017}.
At level 1 of the Purdue Model, where PLCs reside, east-west lateral movement is almost non-existent.

MITRE's ATT\&CK for ICS \cite{MitreICSAttack} lists seven techniques associated with the lateral movement tactic; Default Credentials, Exploitation of Remote Services, Hardcoded Credentials, Lateral Tool Transfer, Program Download, Remote Services, and Valid Accounts~\cite{MitreICSLateral}. 
As with the aforementioned historic attacks, the focus is placed on moving laterally to a PLC, not from or through it.

Prior research into lateral movement at level 1 of the Purdue Model is as scarce as evidence of it in historic OT attacks \cite{Lopez2024}. 
In 2023, Wetzels addressed this gap with a deep-dive into the area, coining the term `deep lateral movement' \cite{Wetzels2023}.
The two primary motivations cited for deep lateral movement are to cross lateral perimeters at level 1 and to gain granular control over nested PLCs deeper in the physical process.
The report highlights the lack of prior work, introduces TTPs for achieving deep lateral movement, and presents a proof-of-concept scenario leveraging vulnerabilities disclosed by Forescout shortly before publication.
Although Wetzels outlines TTPs that align with LOTP and touches on them in the scenario, the work is largely focused on exploiting vulnerabilities rather than leveraging existing functionality.

There is no shortage of LOTP TTPs leveraged in historic OT attacks. 
Perhaps the most infamous example is Stuxnet, which not only exploited zero-days for initial access, but ultimately used legitimate Step7 communication channels and block downloads to modify PLC logic without detection \cite{falliere2011w32,langner2011stuxnet}. 
Similar observations have been made in other OT-targeted malware, where adversaries favoured targeting legitimate OT functionality once footholds were established. 

In the research literature, LOTP work has primarily focused on injecting logic-based malware through legitimate code-download mechanisms. 
Examples include `ladder logic bombs', where malicious behaviour is embedded directly in IEC~61131-3 logic \cite{govil2017ladder}, PLC-resident self-propagating worms such as PLC-Blaster \cite{robertson2016plc}, and extortion techniques like Dead Man's PLC, which orchestrates a ransom mechanism purely from controller-native logic \cite{Derbyshire2024}. 
PLCs have even been tricked into seemingly legitimate communications with rogue engineering stations, as demonstrated in the Rogue7 attacks on Siemens controllers \cite{rogue7}. 

Process comprehension at a distance (PCaaD) edged closer to deep lateral movement via LOTP by targeting the PLCs directly rather than through configuration downloads \cite{Greenpcaad2021}.
It demonstrates how standardised PLC library functions and predictable memory structures can be leveraged to support remote process comprehension \cite{green2017significance}, precise variable manipulation, and in-memory command-and-control.
PCaaD illustrates how LOTP approaches can be extended beyond logic injection into full adversarial orchestration.

In summary, deep lateral movement facilitates crossing lateral perimeters in level 1 and controlling nested PLCs.
Doing this via LOTP would provide stealth and longevity due to challenges in detecting and patching legitimate functionality.

%% file: 3Background/3background.tex
\section{Background}
\label{background}
To support understanding of the technical proof-of-concept and its broader risks, this section outlines the key PLC concepts used and the threat models considered for deep lateral movement.

\subsection{PLC Concepts}
\label{plcconcepts}
PLCs are the heart of the modern OT environment, providing the primary functionality to monitor, control, and automate industrial processes. 
While each PLC's deployment is bespoke to its environment, standardisation is retained in a number of key areas through alignment to BSI/IEC standard 61131-3:2013~\cite{BritishStandardsInstitute2013}. 
This standardisation includes support for five industrial programming languages used to write control logic (Ladder Diagram, Function Block Diagram, Sequential Function Chart, Instruction List, and Structured Text), and the concept of Program Organisation Units (POUs). 
The following definitions use and expand on BSI/IEC 61131-3:13 terminology to provide a generalised model of POUs that are vendor-agnostic~\cite{Greenpcaad2021}:

\begin{itemize}
	\itemsep0em
	\item \textbf{Programs}: Are the highest level of organisational unit. 
	They control and enable responses to cyclic, time-based, or interrupt-driven events during program execution.
	They are composed of specific instructions but also Function Blocks and Functions.
	\item \textbf{Function Blocks (FB)}: Contain code that stores their values permanently in memory, remaining available post Function Block execution.
	\item \textbf{Functions}: Provide discrete common functionality, for example, ADD or SQRT. 
	Function POUs can use global variables to permanently store data, but do not have their own dedicated memory (i.e., local variables).
	\item \textbf{Variable Blocks (VB)}: Store program data and can be global (gVB) or local (fVB). 
	The latter are associated with Function Blocks to provide long-term data storage. 
	The VB is an addition to the POU model as defined by BSI/IEC 61131-3:13. 
	This standard describes the use of variables in a general sense, with limited ties to their storage.
\end{itemize}

Custom control logic can be written in Programs, FBs, and Function POUs through use of the five industrial programming languages, with Functions and FBs supporting code reuse spanning multiple deployments~\cite{Jacinto2017}. 
Furthermore, PLC vendors provide a range of library Functions and FBs that contain pre-written control logic to cover a range of themes, from simple arithmetic to complex data communication exchanges.
These library Functions and FBs simplify the code writing process and so their use is common practice~\cite{Ljungkrantz2007}.

\subsection{Threat Scenarios}
\label{threatscenarios}

Two motivations to conduct deep lateral movement in an OT environment were highlighted in Section \ref{related}.
Those are to cross lateral perimeters between embedded devices in level 1 of the Purdue Model and to reach nested devices that are not otherwise reachable.
These motivations challenge existing paradigms of perimeters and network segregation, such as the concept of `zones and conduits' promoted by ISA/IEC 62443 \cite{ISA62443}.

The following four scenarios illustrate the threat posed by deep lateral movement and the capabilities afforded by the novel technique introduced in this paper.
Across all scenarios, the adversary is assumed to have protocol-level network access to PLC 1 only, with no access to the engineering workstation, project files, or PLC program code; the technique abuses only native PLC communications and functionality.
Scenarios 1--3 target PLC 2 via PLC 1, whereas the final scenario targets PLC 3 via multiple intermediary hops.

\subsubsection{Non-Routable Targeting}
A PLC may be considered safe because routing to it is deemed inaccessible for adversaries.
This could be because the target PLC is restricted to only communicating with operationally necessary PLCs via network architecture.
It could also simply be that an intermediary PLC is dual homed and the target PLC is otherwise not networked.

\begin{figure}[H]
    \centering
    \includegraphics[width=0.8\linewidth]{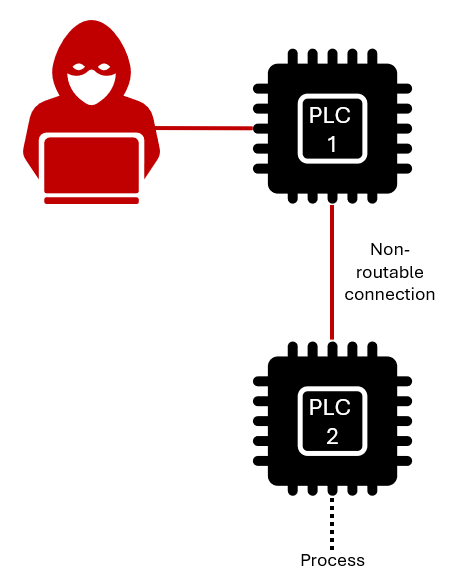}
    \vspace{-10pt}
    \caption{Adversary reaching non-routable PLC 2 through intermediary PLC 1}
	\label{fig:nonroutable}
    \vspace{-10pt}
\end{figure}

Figure \ref{fig:nonroutable} depicts a scenario whereby the adversary wants to reach PLC 2 but it is only routable through PLC 1.
Deep lateral movement through PLC 1 facilitates access to PLC 2 for the adversary, allowing them to further their attack.

\subsubsection{IP to Serial}
Similar to non-routable targeting, a PLC may be considered safe if it is only accessible through a serial connection to another PLC.
There are many technical and architectural reasons why a serial connection may be considered.
One such reason could be the bad practice separation of the basic process control system (BPCS) and the safety instrumented system (SIS), predicated on the assumption that lateral movement over serial is not possible.

\begin{figure}[H]
    \centering
    \includegraphics[width=0.8\linewidth]{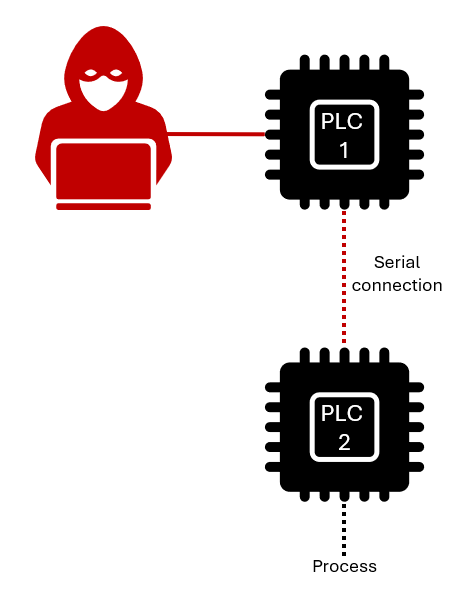}
    \vspace{-10pt}
    \caption{Adversary reaching PLC 2 over serial through intermediary PLC 1}
	\label{fig:serial}
    \vspace{-10pt}
\end{figure}

Figure \ref{fig:serial} depicts a scenario whereby the adversary cannot reach PLC 2 as it is only connected to PLC 1 via serial.
However, the novel deep lateral movement technique introduced in this paper uses a LOTP approach, meaning it can leverage existing communications of intermediary PLCs.

\subsubsection{Site-to-Site}
If an organisation has OT sites dispersed over a large geographic area, they will likely need to facilitate communications between them.
This is often facilitated by private wide area networks (WANs) and will allow their operational assets, such as PLCs, to communicate telemetry and other operational traffic between sites.
This may be considered secure if such sites are not publicly accessible via external connectivity and only allow communications from a restricted list of assets.

\begin{figure}[H]
    \centering
    \includegraphics[width=0.8\linewidth]{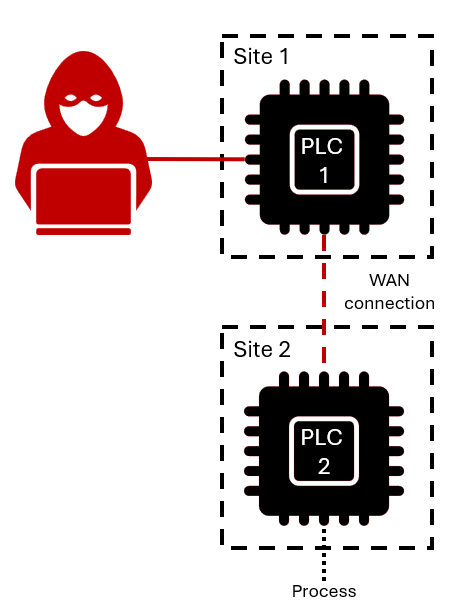}
    \vspace{-10pt}
    \caption{Adversary reaching PLC 2 in site 2 over WAN through intermediary PLC 1}
	\label{fig:WAN}
    \vspace{-10pt}
\end{figure}

Figure \ref{fig:WAN} depicts a scenario whereby the adversary cannot reach site 2 or PLC 2 as access is restricted to operationally necessary devices.
However, because the technique presented in this paper utilises existing communications channels, it facilitates that access via PLC 1.

\subsubsection{Multiple hops}
\begin{figure}[H]
    \centering
    \includegraphics[width=\linewidth]{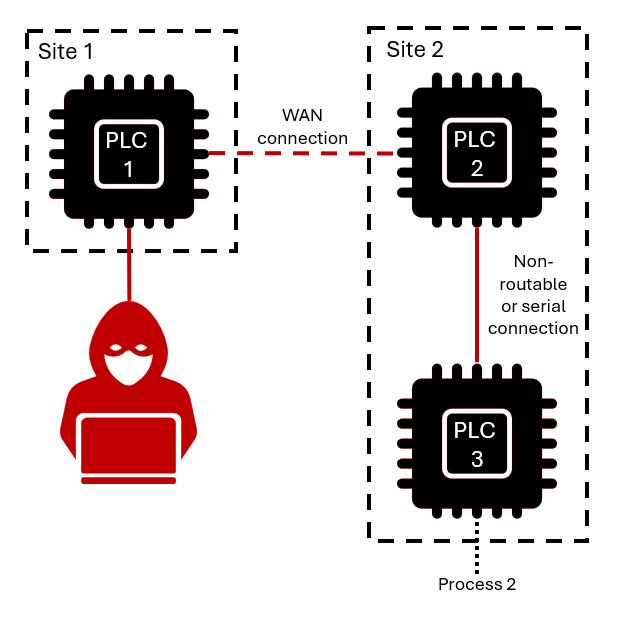}
    \vspace{-10pt}
    \caption{Adversary reaching PLC 3 via hopping through multiple intermediary PLCs}
	\label{fig:hop}
    \vspace{-10pt}
\end{figure}

Finally, the technique presented in this paper is repeatable, which means it is possible to traverse multiple scenarios as above in the same chain of deep lateral movement.
Figure \ref{fig:hop} depicts a simple example of multi-hop deep lateral movement.

%% file: 4Implement/4implement.tex
\section{Practical Implementation}
\label{practical}
\vspace{-10pt}

In search of a viable approach by which deep lateral movement can be realised through the adoption of LOTP techniques, the Siemens PLC ecosystem, more specifically the S7-300 series, was explored.
The S7-300 series has established itself as a widely adopted solution across various sectors since its inception in the mid-1990s, and will continue to receive support from Siemens until 2033~\cite{S7300Support}. 
This enduring relevance and proliferation highlight the significance of S7-300 devices in industrial automation environments.

While exploring the broad range of S7-300 device capabilities, a number of options towards the practical execution of deep lateral movement were identified.
The remainder of this section will focus on one specific technique that can be executed without the need for changes to primary PLC code blocks (i.e., programs, FBs, or Functions), instead relying solely on the manipulation of fVB data through the use of read/write requests.
This technique not only delivers against the primary requirement of deep lateral movement through the use of LOTP, but would be challenging to detect/prevent, and acts as a catalyst towards the reconsideration of traditional OT defensive practices.

\vspace{-10pt}

\subsection{Siemens S7-300 Ecosystem}
Before stepping into a practical breakdown of each attack stage, it is important to understand two key characteristics of the Siemens S7-300 series. 
These are discussed across the following subsections, and are leveraged during the attack (see Section~\ref{attackstages}).

\vspace{-10pt}

\subsubsection{S7comm Protocol}
\label{s7comm}
The Siemens S7comm protocol facilitates network-level interactions with S7-300 PLCs (among others).
These interactions include, but are not limited to, the extraction of diagnostic information and metadata, configuration changes, and operational variable state monitoring and control. 
Operational variable state monitoring and control uses S7comm read/write requests, and represents the most common form of PLC interactions. 
Read/write requests are leveraged by a range of devices, from supervisory control and data acquisition (SCADA) systems, to human-machine interfaces (HMIs), and remote telemetry units (RTUs). 
The primary purpose is to afford visibility and control over operational process conditions to a range of end users (e.g., site operators and integrated control centres).
This represents a significant challenge when implementing security controls, as any errors in their deployment could restrict operational process visibility and control, and therefore safety.

\subsubsection{Siemens Library Functions}
\label{libraryfunctions}
Across all Siemens PLCs, a broad range of library Functions and FBs are made available to programmers via associated programming software (e.g., TIA Portal). 
These provide programmers with ready-made code blocks that can be used to speed up the programming process \cite{Derbyshire2023}.
While some are simplistic in nature, supporting basic calculations, others are more complex, such as those used to facilitate PLC-to-PLC communications.
These FBs and associated lVBs (referred to as Data Blocks, or ``DBs'', in the Siemens ecosystem), are widely used and trusted by PLC programmers, despite known vulnerabilities \cite{Maesschalck2023}.

\subsection{Attack Stages}
\label{attackstages}
The following subsections provide an attack breakdown aligned to key stages and the threat scenarios in Section~\ref{threatscenarios}.

\subsubsection{Identifying PUT/GET FBs}
\label{identify}
Siemens provide a range of library Functions and FBs within their PLC programming software, including PUT and GET FBs~\cite{PUTGET}, which are targeted as part of the attack. 
These two FBs allow a PLC to send (PUT), or collect (GET), data to/from neighbouring PLCs.
PUT/GET communication can be via IP (LAN/WAN) or serial links, as described in the four scenarios in Section \ref{threatscenarios}.

\begin{figure*}[t]
    \centering
    \includegraphics[width=\textwidth]{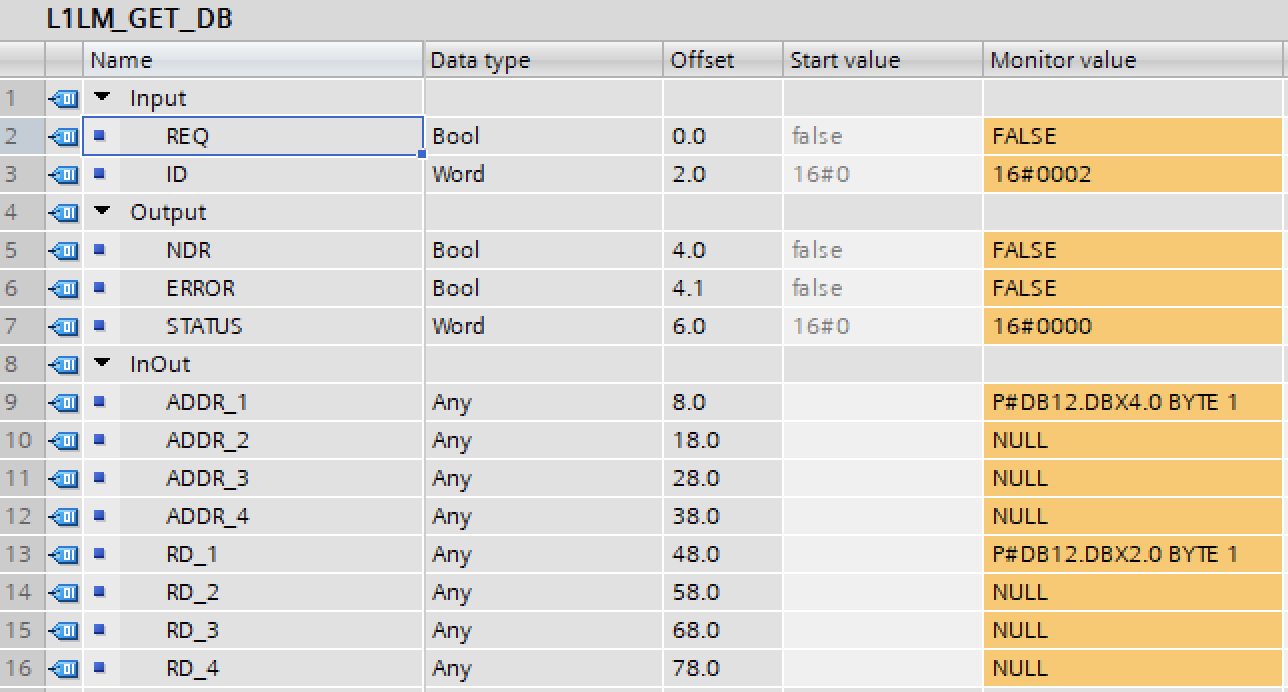}
    \caption{Siemens GET DB}
	\label{fig:getdb}
    \vspace{-10pt}
\end{figure*}

Our previous work focused on the identification of Siemens library FBs via three enumeration techniques (Metadata, Bulk Transfer, and Memory Address Interrogation)~\cite{Greenpcaad2021}. 
The Memory Address Interrogation technique is adopted here, as it requires the use of S7comm read requests alone. 
This approach facilitates the enumeration of PUT and GET FBs with 100\% accuracy, whilst also decreasing the probability of detection and access restrictions. 
Enumeration is achieved via an analysis of PUT and GET lVBs (DBs), based on their size, known values, variable usage, and data type features. 
Figure~\ref{fig:getdb} depicts part of the GET DB. 
Using the Memory Address Interrogation technique, the following steps must be followed:

\begin{enumerate}
    \item Using read requests, check the size of the DB (i.e., starting at byte 0, read one byte at a time until the end of the DB is reached).
    \item Compare the byte count with the GET DB size (600). If they match, move on to 3.
    \item Read all GET DB memory offsets where no data is populated (i.e., unused memory). In Figure~\ref{fig:getdb}, these can be observed at byte offsets 1 and 5. However, several more exist across the entire DB.
    \item Where all areas of memory checked above return a 0 value, the GET FB has been identified.
    \item Additional checks may be undertaken if desired, such as decoding pointer addresses at byte offset 8.
\end{enumerate}

\subsubsection{Check PUT/GET Usage}
PUT and GET allow for the configuration of up to four areas of memory to be sent (PUT) or collected (GET). 
In Figure~\ref{fig:getdb}, this can be seen with the four variables named ADDR\_1, ADDR\_2, ADDR\_3, and ADDR\_4. 
These variables represent the four DB addresses in the target PLC (PLC 2 in all figures in Section~\ref{threatscenarios}) that data can be collected from.
Their configuration is made up of a pointer address, denoted by P\#, followed by the length of data to be collected. 
Data collected from these addresses in the target PLC is then stored in the corresponding variables named RD\_1, RD\_2, RD\_3, and RD\_4.
As with the ADDR variables, these are made up of a pointer address followed by the length of data to be stored. 
To summarise, one byte of data from the target PLC at address DB12.DBX4.0 (ADDR\_1) will be stored in the local PLC address DB12.DBX2.0 (RD\_1).
The configuration of the target PLC sits behind the connection ID (``16\#002'' in Figure~\ref{fig:getdb}), manipulation of which is out of scope.

To understand existing PUT/GET usage, all ADDR values must be read and decoded using S7comm read requests (i.e., convert binary states into pointer and data length values).
Where no data is returned across ADDR address space (i.e., byte offsets 8 to 47 inclusive), it can be considered unused. 
Where one or more of the four PUT/GET configuration parameters is available, the attack can be executed.

\begin{figure*}[t]
    \centering
    \includegraphics[width=\textwidth]{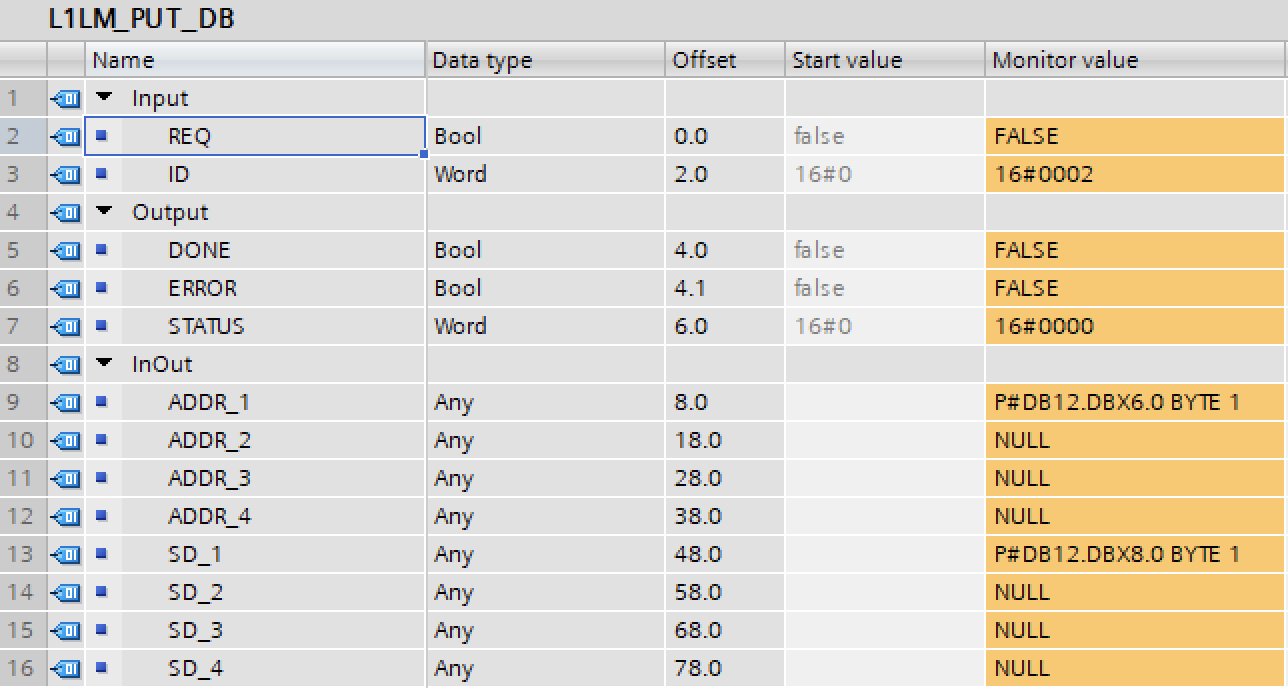}
    \caption{Siemens PUT DB}
	\label{fig:putdb}
    \vspace{-10pt}
\end{figure*}

\subsubsection{Configure a Spare PUT/GET Request}
Taking Figure~\ref{fig:getdb} as an example, it can be seen that ADDR\_2, ADDR\_3, and ADDR\_4 are all available for use. 
Therefore, to collect data from the target PLC, a pointer address and data length must be configured in ADDR\_2, along with a corresponding location to store the collected data in RD\_2. 
During the identification of PUT/GET FBs in Section~\ref{identify}, the memory structure of PUT/GET DBs was leveraged, more specifically areas of unused memory. 
These areas of unused memory offer a location in which data can be stored for collection (GET) or transmission (PUT). 
Therefore, in configuring ADDR\_2 to ``P\#DB1.DBX0.0 INT 1'' and RD\_2 to ``P\#DB100.DBX1.0 INT 1'', upon execution of this GET function, the byte of data stored at address DB1.DBX0.0 within the target PLC will be collected and stored in the local PLC address DB100.DBX1.0.
The same principle would apply for a PUT request. Using Figure~\ref{fig:putdb} as an example, in configuring ADDR\_2 to ``P\#DB1.DBX0.0 INT 1'' and SD\_2 to ``P\#DB101.DBX1.0 INT 1'', upon execution of this PUT function, the byte of data stored at address DB101.DBX1.0 within the local PLC will be sent to address DB1.DBX0.0 in the target PLC.
In the PUT example, the value to be sent to the target PLC must be set at address DB101.DBX1.0 prior to configuring ADDR\_2 and SD\_2, otherwise the function would send a value of 0. 

The configuration of all parameters discussed here is achieved through the use of S7comm write requests alone.

\subsubsection{Monitor Execution State}
Variable ``REQ'' in Figures~\ref{fig:getdb} and~\ref{fig:putdb} shows the current execution state of GET and PUT functions, respectively.
This is a boolean, where 0 represents not executing, and 1 represents executing. 
Therefore, once a request has been configured to send or collect data, monitoring of the REQ variable forms a key next step. 
This can be achieved through the use of S7comm read requests targeting the REQ variable at byte offset 0.

\subsubsection{Remove or Reconfigure}
Once the GET FB has executed, another S7comm read request is required to view the collected data. 
Following the earlier example, this would involve constructing a request targeting one byte of data from DB100.DBX1.0 (the unused area of memory configured to store the collected data). 
ADDR\_2 could then be reconfigured to collect more data, or be set back to 0 along with RD\_2.

Where a PUT FB has executed, it can then be reconfigured to send either a different value (i.e., setting a new value in DB100.DBX1.0 from the earlier example), send data to a different area of memory in the target PLC through the modification of ADDR\_2, or both. 
Alternatively, as with the GET example, all values could be set back to 0.

\subsubsection{Summary}
Where the directly accessible PLC has both PUT and GET functions configured for the same target PLC (denoted by the aforementioned connection ID), a comprehensive attack is possible. 
Using the Memory Address Interrogation technique from our previous work, is not only of value in the identification of PUT/GET FBs on a PLC with a direct connection, but also the target PLC. 
The Memory Address Interrogation technique from our previous work is not only of value in the identification of PUT/GET FBs on a PLC with a direct connection, but also the target PLC.
Using the GET function, it is possible to enumerate Siemens Library FBs on the target PLC too. 
This allows not only for the targeted extraction of operational data (e.g., usernames/password~\cite{Greenpcaad2021}), but also the targeted manipulation of operational data (e.g., communication interface settings~\cite{Greenpcaad2021}) through use of the corresponding PUT function.
In addition, where the GET function and Memory Address Interrogation technique identify the existence of PUT/GET functions in the target PLC, this can be leveraged to pivot further into an environment, where the initial target PLC becomes a secondary jump-off point to reach additional PLCs, as per Figure~\ref{fig:hop}.

%% file: 5Conclusion/5conclusion.tex
\section{Conclusion and Future Work}
\label{conclude}

This paper began by articulating the relative scarcity of research into both LOTP and deep lateral movement within OT.
Both approaches pose significant threats to OT environments in isolation, and more so when combined, as illustrated by the threat scenarios presented in Section~\ref{background}. 
This highlights the importance of exploring their intersection.

The practical implementation described in Section \ref{practical} demonstrates the feasibility of deep lateral movement using a LOTP approach.
Not only does this afford an adversary the capability to traverse laterally between PLCs, it does so in a stealthy manner while leveraging PLC-native functionality.
PLCs previously assumed to be isolated, whether on remote sites via WAN, confined to serial-only networks, or hidden behind intermediary PLCs, are no longer beyond reach. 
By co-opting the very function blocks intended to connect them, adversaries can undermine assumptions of safe network segregation.

The broader implications of the technique introduced here fundamentally challenge the existing approaches to security boundaries and how we control the flow of data between them.
Best-practice concepts, such as ISA/IEC 62443's zones and conduits, must take deep lateral movement into consideration and provide appropriate guidance.
Otherwise, misinterpretation of existing concepts may lead to zones being considered to be isolated when they are not.

While the practical implementation focuses on the Siemens ecosystem, that does not mean it is a hard limitation.
Rather, the technique moves a theoretical concept into demonstrable practice.
Further work should explore the ecosystems of Siemens and other vendors more deeply to understand the prevalence of the technique identified in this paper, as well as alternative techniques that achieve the same effect.